\documentclass[aip,graphicx,amsmath,amssymb,reprint]{revtex4-2}

\usepackage{epsfig}
\usepackage{epstopdf}
\usepackage{graphicx}
\usepackage{dcolumn}
\usepackage{bm}
\usepackage[utf8]{inputenc}
\usepackage[T1]{fontenc}
\usepackage{mathptmx}
\usepackage{etoolbox}

\draft 

\begin{document}


\title{Free-running 4H-SiC single-photon detector with ultralow afterpulse probability at 266 nm} 



\author{Chao Yu}
\affiliation{Hefei National Research Center for Physical Sciences at the Microscale and School of Physical Sciences, University of Science and Technology of China, Hefei 230026, China}
\affiliation{CAS Center for Excellence in Quantum Information and Quantum Physics, University of Science and Technology of China, Hefei 230026, China}

\author{Tianyi Li}
\affiliation{School of Electronic Science and Engineering, Nanjing University, Nanjing 210093, China}

\author{Xian-Song Zhao}
\affiliation{Hefei National Research Center for Physical Sciences at the Microscale and School of Physical Sciences, University of Science and Technology of China, Hefei 230026, China}
\affiliation{CAS Center for Excellence in Quantum Information and Quantum Physics, University of Science and Technology of China, Hefei 230026, China}

\author{Hai Lu}
\email[]{hailu@nju.edu.cn}
\affiliation{School of Electronic Science and Engineering, Nanjing University, Nanjing 210093, China}

\author{Rong Zhang}
\affiliation{School of Electronic Science and Engineering, Nanjing University, Nanjing 210093, China}

\author{Feihu Xu}
\affiliation{Hefei National Research Center for Physical Sciences at the Microscale and School of Physical Sciences, University of Science and Technology of China, Hefei 230026, China}
\affiliation{CAS Center for Excellence in Quantum Information and Quantum Physics, University of Science and Technology of China, Hefei 230026, China}
\affiliation{Hefei National Laboratory, University of Science and Technology of China, Hefei 230088, China}

\author{Jun Zhang}
\email[]{zhangjun@ustc.edu.cn}
\affiliation{Hefei National Research Center for Physical Sciences at the Microscale and School of Physical Sciences, University of Science and Technology of China, Hefei 230026, China}
\affiliation{CAS Center for Excellence in Quantum Information and Quantum Physics, University of Science and Technology of China, Hefei 230026, China}
\affiliation{Hefei National Laboratory, University of Science and Technology of China, Hefei 230088, China}

\author{Jian-Wei Pan}
\affiliation{Hefei National Research Center for Physical Sciences at the Microscale and School of Physical Sciences, University of Science and Technology of China, Hefei 230026, China}
\affiliation{CAS Center for Excellence in Quantum Information and Quantum Physics, University of Science and Technology of China, Hefei 230026, China}
\affiliation{Hefei National Laboratory, University of Science and Technology of China, Hefei 230088, China}

\date{\today}

\begin{abstract}
Ultraviolet single-photon detector (UVSPD) provides a key tool for the applications requiring ultraweak light detection in the wavelength band. Here, we report a 4H-SiC single-photon avalanche diode (SPAD) based free-running UVSPD with ultralow afterpulse probability. We design and fabricate the 4H-SiC SPAD with a beveled mesa structure, which exhibits the characteristic of ultralow dark current. We further develop a readout circuit of passive quenching and active reset with tunable hold-off time setting to considerably suppress the afterpulsing effect. The nonuniformity of photon detection efficiency (PDE) across the SPAD active area with a diameter of $\sim$ 180 $\mu$m is investigated for performance optimization. The compact UVSPD is then characterized, exhibiting a typical performance of 10.3\% PDE, 133 kcps dark count rate and 0.3\% afterpulse probability at 266 nm. Such performance indicates that the compact UVSPD could be used for practical ultraviolet photon-counting applications.
\end{abstract}


\maketitle 

\section{Introduction}

Weak light detection in the ultraviolet (UV) band is widely used for diverse applications such as corona discharge detection~\cite{YYX18}, non-line-of-sight communications~\cite{DEV20}, biological detection~\cite{YKN16}, UV astronomy~\cite{MPU09}, and UV Lidar~\cite{RWR17}. Single-photon detector (SPD) provides a kind of tool with the limit of sensitives for weak light detection. Currently, the primary approach to achieve single-photon detection in the UV band is using commercial available photomultiplier tubes. However, such devices suffer some disadvantages such as requiring vacuum tube and relatively large size. Nevertheless, 4H-SiC single-photon avalanche diodes (SPADs) have been considering as an alternative technology to implement UVSPD~\cite{XXD07,XDH07,ASA09,HLD19,LDH19}, which might be similar to other semiconductor devices of SPADs in the visible band and the infrared band.

Using 4H-SiC SPADs for photon counting was first demonstrated in 2005~\cite{XFX05}. Since then, different junction structures and fabrication technologies have been reported, and therefore the performance of 4H-SiC SPADs has been gradually improved~\cite{AGS05,XHD09,XLH18,LDQ20}. However, two major drawbacks of previous 4H-SiC SPADs have to be overcome to implement practical uses. On one hand, the afterpulsing effect of 4H-SiC SPAD is considerably severe. In the free-running mode, the afterpulse probability ($P_{ap}$) could reach as high as several hundreds of percent~\cite{YYY18}. With double-gate method~\cite{ZIZP15}, further experimental result on $P_{ap}$ distribution shows that once an avalanche occurs the subsequent $P_{ap}$ is roughly 100\% and after several hundred nanoseconds $P_{ap}$ decreases down to 0~\cite{HHL20}, which implies that the lifetime of trapped carrier could be quite short. On the other hand, the distribution of avalanche gain across the active area of 4H-SiC SPAD exists severe nonuniformity, which results in large variations of avalanche amplitudes~\cite{YYY18,FDH15} and thus the nonuniformity of photon detection efficiency (PDE)~\cite{XCH17}.

\begin{figure*}[tbp]
\centering
\includegraphics[width=17 cm]{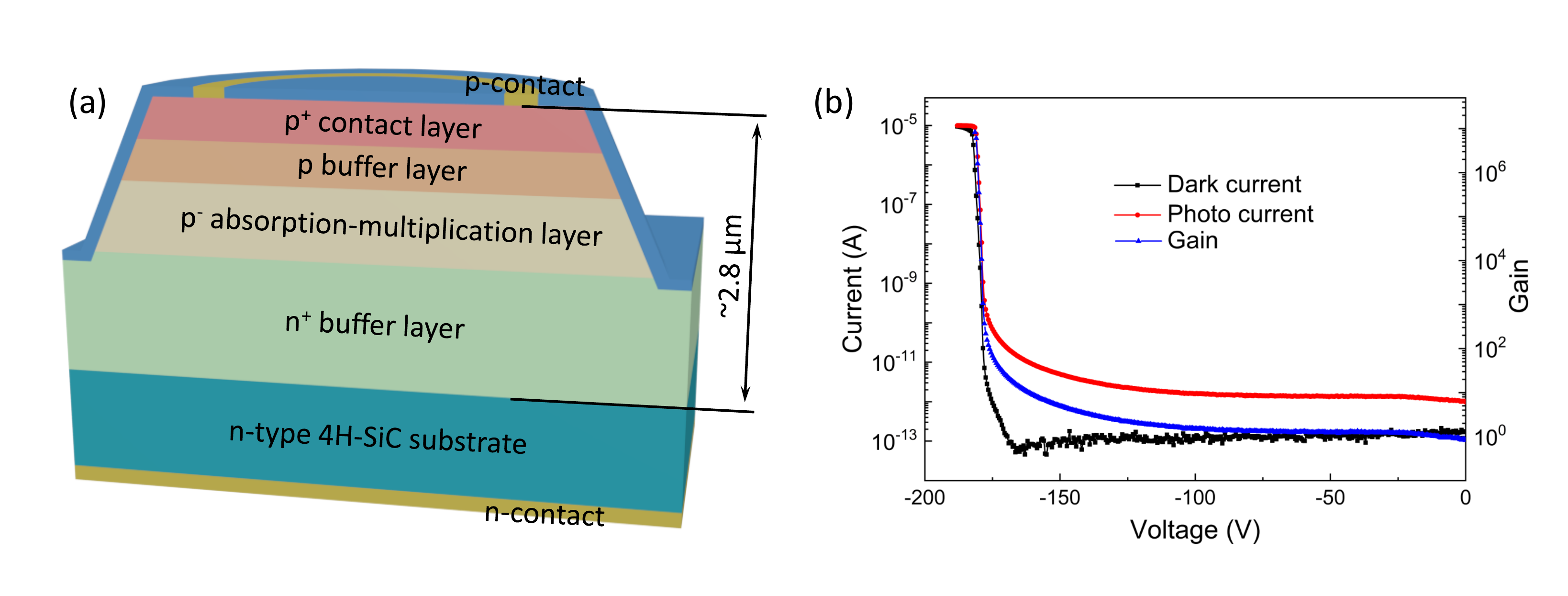}
\caption{(a) Schematic diagram of 4H-SiC SPAD structure with a beveled mesa. (b) Room temperature current-voltage and gain-voltage characteristics of 4H-SiC SPAD.}
\label{fig1}
\end{figure*}

In this paper, we fabricate a 4H-SiC SPAD with optimized structure design, and further implement a compact and practical free-running UVSPD with a dedicated readout circuit of passive quenching and active reset to suppress the afterpulsing effect. By scanning across the active area of SPAD with 266 nm laser, we calibrate the nonuniformity distribution of PDE. Further, we characterize the primary parameters of UVSPD with results of 10.3\% PDE, 133 kcps dark count rate (DCR), and 0.3\% $P_{ap}$ at -40 $^{\circ}$C. We believe that our present UVSPD could be ready for some potential applications despite the fact that the parameters should be further improved.

\section{SPAD DESIGN AND FABRICATION}

The schematic structure of the 4H-SiC SPAD is illustrated in Fig.~\ref{fig1}(a). The SPAD is fabricated on a 4-inch n-type 4H-SiC substrate. Above the substrate, the epitaxy structure from bottom to top includes a heavily-doped $n^{+}$ buffer layer, a $p^{-}$ absorption-multiplication layer, a $p$ buffer layer, and a $p^{+}$ contact layer. Due to the strong absorption of deep UV photons, the top $p^{+}$ contact layer is designed to be thin enough to avoid excess absorption loss. However, ohmic metallization requires a certain thickness, and thin contact layer may induce that alloying spikes reach the high field region and thus result in premature breakdown. Considering the trade-off of the above factors, the thickness of $p^{+}$ contact layer is designed to be 0.1 $\mu$m. Meanwhile, the thickness of the absorption-multiplication layer is designed to be 0.5 $\mu$m, which ensures sufficient carrier acceleration distance while keeping the avalanche breakdown voltage less than $\sim$ 200 V. The total thickness of the epitaxy layers is $\sim$ 2.8 $\mu$m and the diameter of active area is $\sim$ 180 $\mu$m.

When a SPAD is operated under high electric field, electric field crowding existing around the device edge could result in lower breakdown voltage and higher dark current, which degrades the device performance. For our SPAD, beveled mesa termination structure with a small slope angle of $\sim$ 4.5$^{\circ}$ is fabricated using a photoresist reflow technique following the process of inductively coupled plasma etching. During the etching process, surface damage related defects could occur on the sidewall of the mesa, which increases the device dark current and reduces the device reliability in the case of high reverse bias. To eliminate the surface defects induced leakage, after growing and removing a sacrificial thermal oxide layer, the SPAD device is deposited with 1 $\mu$m $SiO_{2}$ using plasma enhanced chemical vapor deposition and further passivated using thermal oxidation. The contact windows are opened by the processes of photolithography and wet chemical etching. The design of n-type and p-type contact layers adopts Ni/Ti/Al/Au multi-metal-layer scheme for preventing oxidization of metals and obtaining a relatively stable ohmic contact. Both contact layers are deposited using E-beam evaporation after the process of rapid thermal annealing at 850$^{\circ}$C in $N_{2}$ environment. Fig.~\ref{fig1}(b) plots the test results of dark current, photo current, and gain as a function of bias voltage in room temperature, from which one can observe that the avalanche breakdown voltage of SPAD reaches $\sim$ 178 V.

\begin{figure*}[htbp]
\includegraphics[width=17 cm]{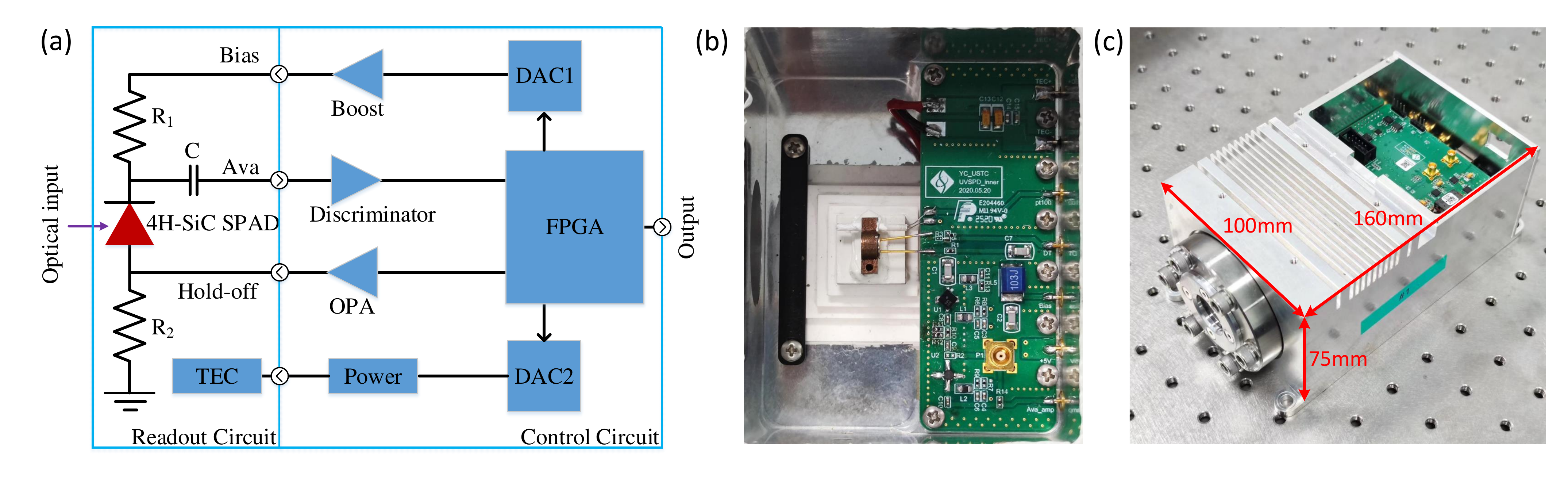}
\caption{(a) Schematic circuit diagram of the 4H-SiC SPD. (b) Photo of the SPAD and the readout circuit. (c) Photo of the whole UVSPD.}
\label{fig2}
\end{figure*}

\begin{figure}[htbp]
\includegraphics[width=8.5 cm]{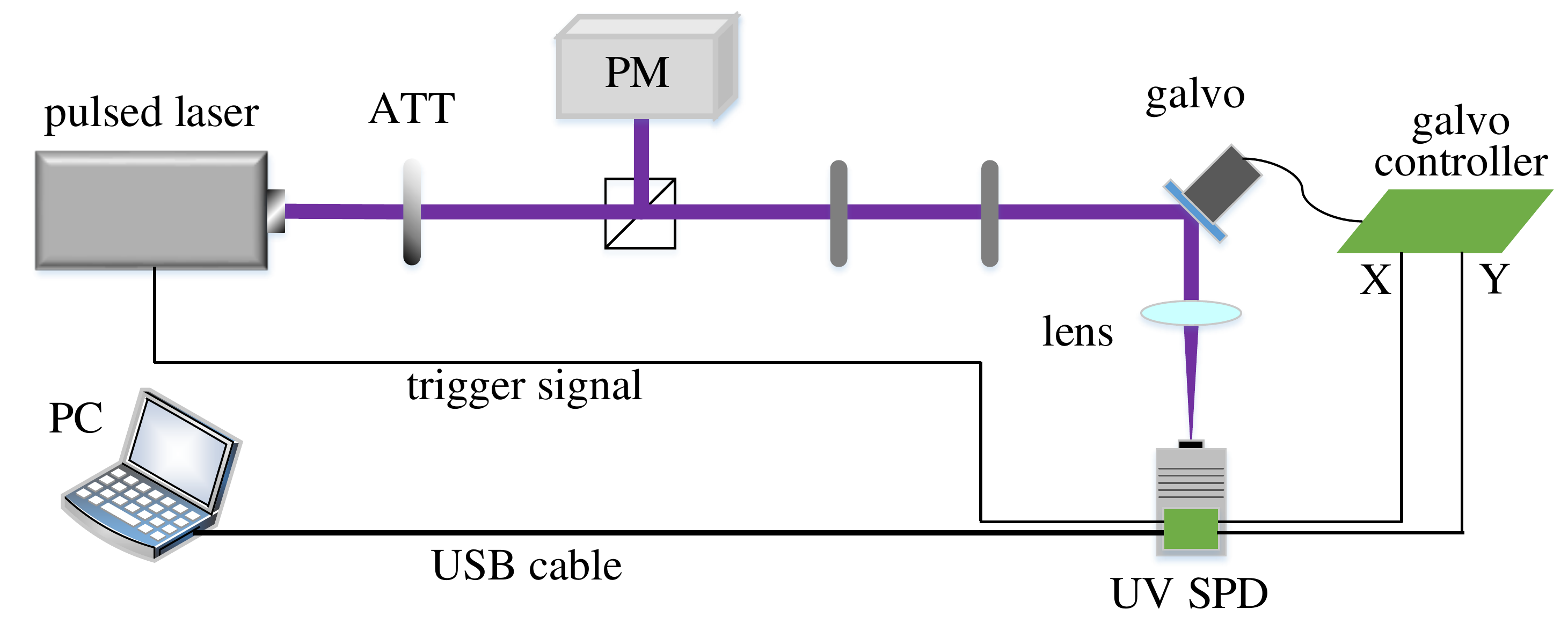}
\caption{Experimental setup for UVSPAD characterization.}
\label{fig3}
\end{figure}

\section{SPD SYSTEM AND CHARACTERIZATION}

Based on the 4H-SiC SPAD, we further implement a free-running UVSPD. The schematic circuit diagram of the SPD is illustrated in Fig.~\ref{fig2}(a), including a readout circuit and a control circuit installed in two cavities, respectively. The electronic signals in two cavities are connected by through-hole micro coaxial connectors. In the readout circuit cavity, the SPAD is fixed on top of a thermoelectric cooler (TEC), and the readout circuit is installed close to the SPAD, as shown in Fig.~\ref{fig2}(b). A 200 k$\Omega$ resistor ($R_{1}$) is connected to the cathode of SPAD for passive quenching, and avalanche signals are alternating current coupled to a discriminator via a 10 pF capacitor. Hold-off signals are connected with the anode of SPAD via a 50 $\Omega$ resistor ($R_{2}$). Once an avalanche signal is discriminated, the hold-off signal is pulled up to 10 V and lasted for a certain hold-off time regulated by a field-programmable gate array (FPGA). Such approach can effectively suppress the afterpulsing effect~\cite{ZIZP15}. When the hold-off signal changes from the high-level to the low-level, the SPAD is actively reset to detect the next photon.

\begin{figure}[htbp]
\includegraphics[width=8.5 cm]{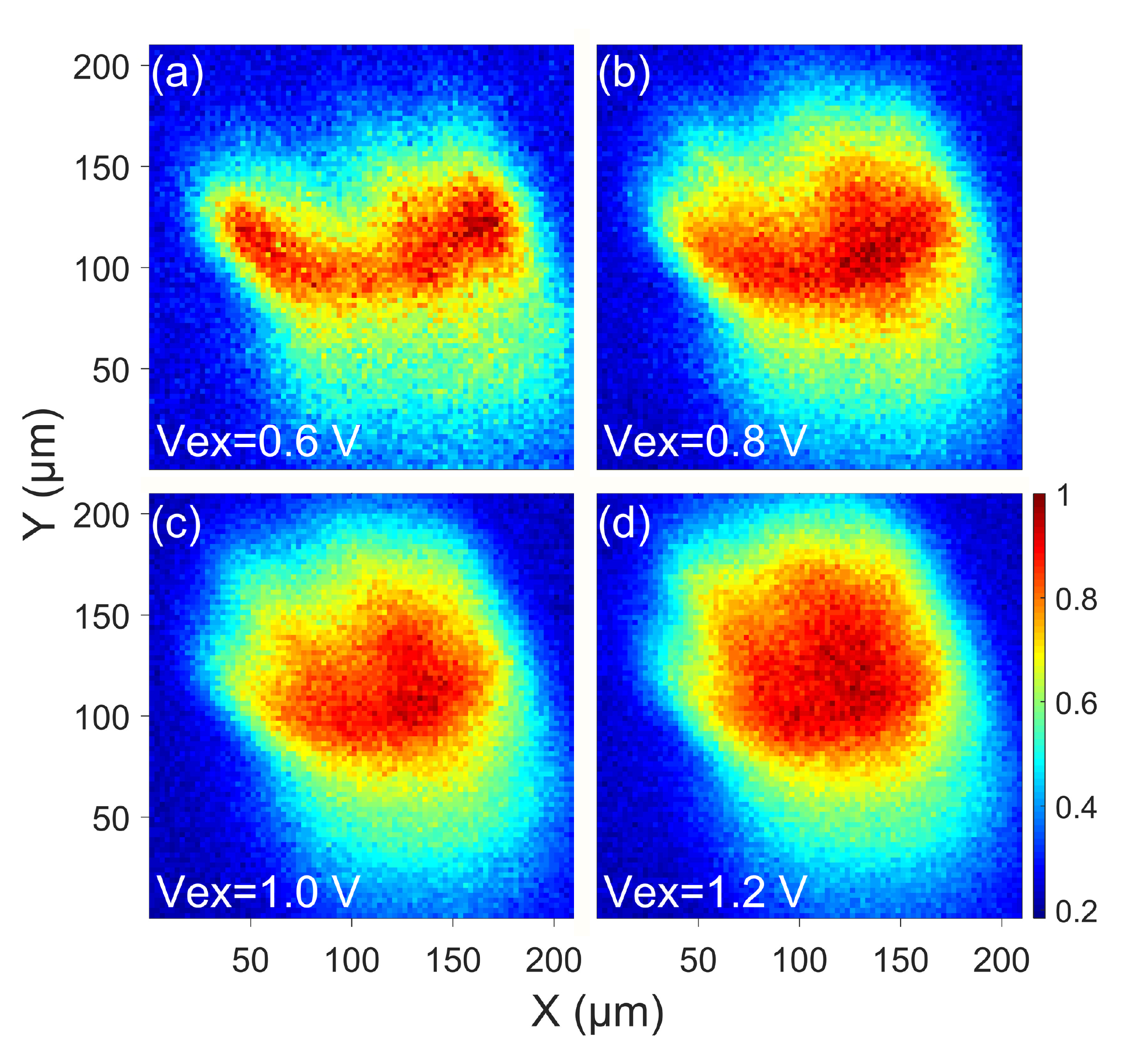}
\caption{Distribution of the normalized photon detection efficiency across the active area of SPAD with excess bias voltages of 0.6 V (a), 0.8 V (b), 1.0 V (c), and 1.2 V (d). The scanning step is 2.5 $\mu$m and the acquisition time is 1 ms at each position point.}
\label{fig4}
\end{figure}

In the control circuit, the FPGA is used to control the whole electronics of UVSPD. A digital-to-analog converter (DAC1) outputs a programmable reference voltage, which is then amplified to be used as bias voltage by a boost circuit module. DAC2 regulates the output power of the TEC driver, which could maintain the SPAD temperature with fluctuation of $\pm$0.1 $^{\circ}$C via a proportional-integral-derivative algorithm inside the FPGA. The output signals from the discriminator are fed into the FPGA. The FPGA monitors the count rate of avalanche signals, and generates hold-off signals amplified by an operational amplifier (OPA), and outputs synchronized detection signals. Meanwhile, a time-to-digital converter (TDC) module with a resolution of 10 ns is programmed inside the FPGA for the time measurement of avalanche signals. The size of the whole UVSPD is 160 mm$\times$100 mm$\times$75 mm, as shown in Fig.~\ref{fig2}(c), and its weight is 1.5 kg.

We then perform the characterization of SPD parameters. Fig.~\ref{fig3} illustrates the experimental setup.
A pulsed laser generates a collimated beam with 100 ps pulse width and 1 mm diameter at 266 nm. After passing through a tunable attenuator (ATT), the laser beam is divided into two parts by a beamsplitter, where one port is monitored by a power meter (PM) and the other port is further attenuated to the single-photon level by attenuators in series. The attenuated beam is deflected by a scanning galvo system, and finally focused to the active area of SPAD by an aspheric lens with a focal length of 79 mm. The beam diameter on the focal plane is estimated to be $\sim$ 27 $\mu$m. The trigger signals to regulate the frequency of the pulsed laser and the driving signals to regulate the deflection angles of galvo are generated by the control circuit of SPD. For convenient operations, the SPD is connected with a personal computer (PC) via a universal serial bus (USB) interface in the control circuit, and a graphical user interface program is developed to easily configurate the settings of SPD and to receive the data from the FPGA.

During the characterization, we first investigate the nonuniformity of PDE by scanning the focused laser beam across the SPAD active area. The distribution results of the normalized PDE in the cases with different excess bias voltages ($V_{ex}$) are plotted in Fig.~\ref{fig4}. The scanning step of the focused laser beam is fixed as 2.5 $\mu$m. The illumination intensity is set to 10 Mphoton/s and the acquisition time is 1 ms at each position point to achieve an appropriate signal-to-noise ratio.
The results in Fig.~\ref{fig4} clearly exhibit that the PDE distribution across the SPAD active area is nonuniform, particularly in the cases with low excess bias voltages. This phenomenon is essentially due to the nonuniformity of the electric field intensity in the multiplication layer. At the positions with weak electric field intensity, the avalanche probability and the avalanche amplitude could be lower than that with high electric field intensity, which results in less detection events given a fixed threshold for the discriminator. In addition, the anisotropic electrical transport caused by off-orientation epitaxial growth may also contribute to the nonuniformity of avalanche gain across the SPAD active area~\cite{LXH19}.

\begin{figure*}[htp]
\includegraphics[width=18 cm]{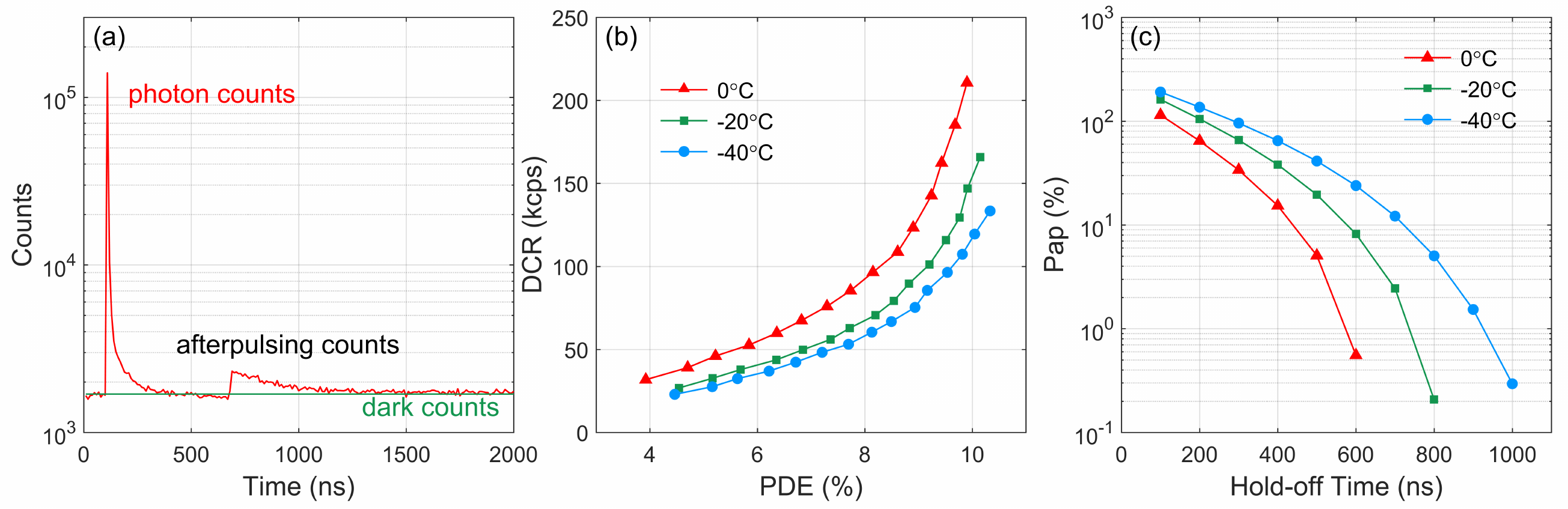}
\caption{(a) Typical count distribution measured by the TDC with 60 s acquisition time. (b) DCR versus PDE at different temperatures. $V_{ex}$ is set from 3 V to 10 V with 0.5 V step. (c) Afterpulse probability versus hold-off time at different temperatures.}
\label{fig5}
\end{figure*}

By fixing the focused laser beam at the position with the highest PDE, we calibrate the primary parameters of the UVSPD including DCR, PDE, and $P_{ap}$, which can be calculated from the TDC data. The repetition frequency of the pulsed laser is set to 50 kHz and its intensity is attenuated to 1 photon/pulse. Fig.~\ref{fig5}(a) shows a typical TDC data with 60 s acquisition time. The first peak corresponds to photon counts of laser pulse ($C_{p}$), and the second small peak after a period of hold-off time corresponds to afterpulsing counts ($C_{ap}$). The afterpulse probability rapidly decreases down to a negligible level within 1 $\mu$s, and the subsequent constant level corresponds to dark counts ($C_{d}$).

From the count distribution,
DCR is calculated by $DCR=C_{d}/(ft_{bin}T_{acc})$, where $f$ is the repetition frequency of laser pulses, $t_{bin}$ is the time resolution of TDC, and $T_{acc}$ is the acquisition time.
Considering the Poisson distribution of photon number, PDE is calculated by~\cite{CMH17}
\begin{equation}
\label{PDE}
PDE=-\frac{1}{\mu}\ln(1-\frac{C_{p}}{fT_{acc}}),
\end{equation}
where $\mu$ is the mean photon number per pulse. The total afterpulse probability is calculated by $P_{ap}=C_{ap}/C_{p}$.

By increasing $V_{ex}$ from 3 V to 10 V with 0.5 V step, Fig.\ref{fig5}(b) plots the trends of DCR versus PDE with 1 $\mu$s hold-off time at different temperatures of 0 $^{\circ}$C, -20 $^{\circ}$C, and -40 $^{\circ}$C. At the same PDE of $\sim$ 10\%, when temperature decreases from 0 $^{\circ}$C to -40 $^{\circ}$C, DCR decreases by $\sim$ 40\%, i.e., from $\sim$ 210 kcps down to $\sim$ 120 kcps.
Empirically, we could conclude that the temperature dependence of DCR for 4H-SiC SPAD is significantly weaker than that of other SPAD devices in the visible band or in the infrared band.
Fig.\ref{fig5}(c) plots the trends of $P_{ap}$ versus hold-off time with $V_{ex}$=10 V at different temperatures, from which one can observe that the decay slope of $P_{ap}$ becomes steeper at higher temperature. At -40 $^{\circ}$C, $P_{ap}$ drops to 5\% with 0.8 $\mu$s hold-off time, and further down to only 0.3\% with 1 $\mu$s hold-off time. We believe that there are some rooms to improve the SPAD performance in the future. On one hand, one can continue to optimize the SPAD device structure design, e.g., exploiting separate absorption, charge and multiplication structure to enhance the detection efficiency. On the other hand, during fabrication one can reduce the density of defects within the active layers to decrease the noise performance of DCR and $P_{ap}$.

\section{CONCLUSION}

In summary, we have designed and fabricated a 4H-SiC SPAD, and have then implemented a compact UVSPD with with ultralow afterpulse probability by developing dedicated electronics including readout circuit and control circuit. After calibrating the nonuniformity of PDE across the SPAD active area, we have characterized the primary parameters of UVSPD with performance of 10.3\% PDE, 133 kcps DCR, and 0.3\% $P_{ap}$ at -40 $^{\circ}$C. The performance and the compactness feature could make the UVSPD suitable for some practical applications.
\section*{acknowledgments}
This work has been supported by the National Key R\&D Program of China (2017YFA0304004); National Natural Science Foundation of China (62175227); Chinese Academy of Sciences; Anhui Initiative in Quantum Information Technologies.

\section*{DATA AVAILABILITY}
Data underlying the results presented in this paper are not publicly available at this time but may be obtained from the authors upon reasonable request.


\end{document}